# Single nanosized graphene/TiO$_x$ multi-shells on TiO$_2$ core via rapid-concomitant reaction pathway on metal oxide/polymer interface

Kunihiko Kato,[1] Yunzi Xin,[1] Sébastien Vaucher[2] and Takashi Shirai *[1]

[1.] Advanced Ceramics Research Center, Nagoya Institute of Technology, Gokiso, Showa-ku, Nagoya, Aichi 466-8555 Japan.

[2.] Swiss Federal Laboratories for Materials Science and Technology (Empa), Feuerwerkerstrasse 39, Thun CH-3602 Switzerland

*Corresponding author.    *E-mail*: shirai@nitech.ac.jp

## Abstract

A novel design has been proposed for a facile and rapid build-up of highly tailorable nanostructured multishells on metal oxide particles (graphene/TiO$_x$@TiO$_2$) under a dry inert atmosphere to maximize the visible-light photocatalytic performance. We also thoroughly and systematically investigated the as-prepared nanostructured particle surface and the mechanisms of the extraordinary in-situ graphene growth on the TiO$_x$@TiO$_2$ nanoparticles achieved by a rapid-concomitant reaction in the metal oxide (TiO$_2$)/polymer (polymethyl methacrylate) interface under microwave irradiation. The as-prepared composite materials are also found to perform a better photocatalytic activity in comparison to the traditional synthesis pathway for degradation of organic pollutants under visible light, associated with the synergetic effects of homo-/hetero-junction on TiO$_2$.

Designing complex functional surfaces with well-controlled nanostructure, specific chemical and physical properties is one of the pivotal challenges faced by material scientists. In this framework, low dimensional carbon (graphene, fullerene, carbon nanotubes, and nanofibers) coating on inorganic semiconductor particles is gaining considerable attention because of the outstanding electric properties, chemical stability, and mechanical properties, not only in the development of advanced functional materials for high-performance electronic materials, lithium-ion batteries, solar cells, etc. [1-3] but also for heterogeneous catalysts including photocatalysts [4]. In the case of titanium oxide ($TiO_2$), one of the most intensively investigated photocatalysts, appropriate post-processing by surface modification for defects engineering greatly affects extending the absorption range to visible or even infrared photons and allows efficient separation of the photo-excited carriers [5].

Recently, the novel methodology for synthesizing $TiO_2$ nanoparticles embedded in graphene sheet has been developed by microwave (MW)-assisted liquid-phase reactions followed by the loading of $TiO_2$ nanoparticles or precursors into the solution with the exfoliated/dispersed graphene oxide (GO) nanoparticles [6-8]. Besides, another research group has reported the formation of carbon-coated Magnéli-phase $Ti_nO_{2n-1}$ nanorods from the $TiO_2$ nanorods dispersed in PVP aqueous solution via carbothermal reduction under MW irradiation [9]. However, there is no study for simultaneous $TiO_2$ surface modification and high-quality graphene coating to maximize the visible-light photocatalytic performance.

Herein, we present a novel and facile process to simultaneously achieve rapid build-up of multi-shelled structure on $TiO_2$ (graphene/$TiO_x$@$TiO_2$) under a dry inert atmosphere. We first report that the extraordinary in-situ graphene growth on the $TiO_x@TiO_2$ nanoparticles results from a rapid concomitant reaction assisted by a MW electric field on the organic polymer (polymethyl methacrylate; PMMA) / metal oxide ($TiO_2$) interface. Furthermore, the photocatalyst with homo-/hetero-junction shows significant enhancement by the synergetic effects on the photo-degradation of Rhodamine B (RhB) under visible light compared to $TiO_2$/nanocarbon composites through a synthesis using resistive furnace heating. Furthermore, the difference in the formation mechanism between the two methods is discussed in depth.

The representative morphology of the composite particles obtained by MW-assisted reaction is shown by HR-TEM image with SAED pattern in Fig. 1a-b (The TEM images of another area are shown in Fig. S4). It reveals the formation of a continuous 5nm thick shell (lighter grey) on the surface of $TiO_2$ particles, which is identified as graphene by the resulting 3.4 Å d-spacing and electron diffraction patterns of graphite (002) [10]. SEM images confirm the disappearance of PMMA particles after the MW-assisted reaction (Fig. S5). Conversely, an additional black contrasted area is also observed at the interface between the surface graphene shell and the $TiO_2$ core. This area is assigned to $TiO_x$ as a stark black contrast is commonly observed for deficient $TiO_2$ [11]. Thus, we assume for the as-prepared $TiO_2$ particles a double-shell structure as illustrated in Fig. 1c. In addition to a colour change from white to black (insert of Fig. 1d), the rapid MW-assisted reaction resulted in a compelling improvement in the light absorption capacity in the UV to visible range and the NIR region. Examining the long-range structural order, neither $TiO_2$ phase transitions nor impurity phases could be detected by XRD (Fig. 2a). However, the presence of graphene on the TiO2 surfaces is clearly evident from the Raman spectra in Fig. 2b, as the peaks associated with the G-band (1335 cm$^{-1}$) and D-band (1590 cm$^{-1}$) were present. Here, ROC, AD and AC represent the residual organic carbon derived from PMMA, amorphised diamond, and amorphous carbon [12]. Using the D-band and G-band peak intensity ratio ($I_D/I_G$), the degree of graphitisation can be estimated at around 0.82. We believe the plastic-derived graphene layer on ceramics particles possesses a highly ordered graphitic structure even without a metal-catalyst, which is comparative quality in graphene-coated ceramics particles from low molecular weight hydrocarbon via the high-temperature CVD method [13].

The XPS spectra in Fig. 2c-d confirm the formation of TiO$_x$ and graphene on the TiO$_2$ particle surface. In the Ti$_{2p}$ spectra, the three peaks at 458.8, 457.1, and 455.5 eV are assigned to Ti$^{4+}$, Ti$^{3+}$, and Ti$^{2+}$, respectively [14]. In comparison to the surface chemical structure, highly concentrated lower Ti valence species created a shoulder in the range of 455 to 459 eV in the XPS Ti$_{2p}$ spectrum after MW-assisted reaction (MW-500). In addition, the C$_{1s}$ spectrum shows that both the peak intensities of O-C=O (289.0 eV) and C-O (286.8 eV) are significantly reduced (detailed peak assignments from the XPS C$_{1s}$ spectrum are summarised in Fig. S6), whereas the sp$^2$ (283.8 eV) and π-π* satellite (293 eV) signals assigned to graphene structure appear strengthened through the rapid MW-assisted reaction. In this case, the formation of the Ti-C bond at 281.8 eV [15] by the carbon dopant was observed as a minor peak. Fig. S7 shows the XRD patterns of the particles prepared through the conventional heating process. No phase transformations are observed, even though anatase is expected to transform into rutile above 950 °C [16], because carbon on $TiO_2$ surfaces suppresses this phase transformation [13,17]. Raman spectroscopy reveals notable differences between the structure of the graphene shell obtained by MW heating and conventional heating. Fig. S8-9 displays the ratio of ROC peak area by the total peak area plotted as a function of the reaction temperature. The samples produced using the MW heating show a significantly lower level of ROC than the ones made by conventional heating, despite a much lower processing temperature. This finding supports the hypothesis of acceleration of the polymer decomposition through a selective MW absorption. Surprisingly, the $I_D/I_G$ ratio increases at higher processing temperatures in both cases. We expected this trend to be opposing because high-temperature heat treatments usually assist the growth of graphitic crystal structures with well-ordered structures. However, oxides of graphite and graphene are more disordered than pure graphite (graphene), which results in strong D-band peak intensities [18]. This result proves that oxygen abstraction from $TiO_2$ co-occurs with the rapid dissociation of the organic polymer chain by the direct interactions with the MW electric field, ultimately leading to the formation of $TiO_x$ and O atoms incorporated into the graphitic structure. As the graphene shell is dense, only a limited amount of oxygen atoms can be extracted from the anatase, thereby limiting the oxygen vacancies at a low enough level to inhibit the allotropic transformation. Both heating processes do not show a significant change in the Brunner–Emmett–Teller specific surface area, which is calculated from $N_2$ adsorption isotherms (Fig. S10-11). According to the XPS C1s spectra of the samples made by the conventional heating process (Fig. S12), in addition to the detection of O-C=O and C-O attributed to ROC, the substitution of carbon atoms at the oxygen sites on the $TiO_2$ lattice is takes place above 500°C resulting in the Ti-C bonding signal. Moreover, the intensity of the signals in XPS $O_{1s}$ spectra, assigned to OH groups (531.7 eV) and $H_2O$ (533.1 eV) [19], increases at higher temperatures. The XPS O1s spectrum of MW-500 and the summary of the percentage values of $O^{2-}$, $OH^-$, and $H_2O$ are shown to compare between the synthesis by MW and conventional heating (in Fig. S13). Generally, oxygen vacancies induce the dissociation of adsorbed $H_2O$ molecules to form two OH groups [20]. Essentially, defects in the C-doped $TiO_2$ are often associated with oxygen vacancies created due to charge compensation [21]. The $Ti_{2p}$ spectra display a shoulder peak attesting a significant concentration of $Ti^{2+}$. These $O_{1s}$ and $Ti_{2p}$ spectra are consistent with an effective carbon atom doping of $TiO_2$. The plot of C-dopant concentration as a function of $Ti^{2+}$ species (calculated by XPS) shows that only those materials made using MW-500 deviate from the trend line (see Fig. S12d), and indicates that in this latter case, the formation of the $TiO_x$ phase was less dependent on the reaction involving the incorporation of C-dopants within the anatase lattice. Comparison of GC analysis of the molecular effluents collected during the two types of heat treatments is shown in Fig. S13. (a comprehensive list of chemical constituents is available in Table S1 and S2). The GC spectrum for conventional heating is in good agreement with Ref. [22]. The main component produced is ethyl methacrylate monomers likely due to the main-chain cleavage. By contrast, under MW irradiation, PMMA releases low-molecular-weight molecules, with C=C bonds, such as propylene and aromatic rings instead of monomers. Fig. S14 compares the thickness of the graphene shell on the $TiO_x$@$TiO_2$ surface using HR-TEM. In the MW-assisted reaction, the thickness of the graphene can be controlled by varying the reaction temperature from 1 to 5-6 nm, a range out of reach in conventional heating. It seems that specific phenomena occur during the MW heating process, promoting the growth of the graphene shell despite the low-temperature treatments (~500 °C). The graphene shell thickness and the calculated yield values of inorganic carbon are summarised in Table S3. Here, we find the rapid chain cleavage of the organic polymer, achieved by the selective MW absorption, gave higher-rate conversions from the organic polymer that transformed into inorganic carbon compounds, especially those with graphitic structures.

Based on the results and discussion above, we propose the possible formation mechanisms of graphene–$TiO_x$ multi-nanoshell $TiO_2$ via the MW-assisted reaction in four steps (Fig. 3). a) Polymer chains dissociation under the action of the MW electric field leads to active low molecular fragments containing unsaturated C=C bonds and aromatic rings, as precursors of the graphene structure (Reaction 1). b) Simultaneously, surface reduction of $TiO_2$ particle via oxygen abstraction by the radical species from the chain cleavage of the polymer (Reaction 2). c) At this point, the incorporation of O atoms released from $TiO_2$ into the structure of graphene precursors proceeds. The formed defective $TiO_2$ ($TiO_x$) surface, on which small molecular fragments adsorbs predominantly at defect sites associated with oxygen vacancies and $Ti^{3+}$ species, promotes dissociative adsorption [23] (Reaction 3). The activation energy for dissociative adsorption of carbon precursor on the substrate surface is usually the rate-limiting reaction step in the growth of graphene [24]. Moreover, nucleation of graphene preferentially occurs at the point defects [25], such as oxygen vacancies and $Ti^{3+}$ species. d) Finally, the synergetic contribution of the $TiO_x$ defects as catalytic sites along with the selective MW absorptivity to the intermediate compounds, accelerate their conversion into oxygen-containing graphene (Reaction 4). By contrast, the particles prepared by conventional heating follow a completely different chemical pathway. Milder reaction conditions with slower polymer decomposition rates generate ethyl methacrylate as a monomer via the main-chain cleavage. Extended chemical contact between surface $TiO_2$ and hydrocarbon at high temperatures favours the formation of C-doped $TiO_x$ similar to carbothermal reduction [26], accompanied by the incorporation of O into the graphitic carbon structure (see Fig. S15).

Fig. 4a shows the superior photocatalytic performance of multi-shelled $TiO_2$ particles for the photo-degradation of RhB under visible-light irradiation. For both raw and coated $TiO_2$, the kinetics of the photocatalytic degradation follows a first-order law, expressed as $\ln(C_0/C) = kt$ [27], where t stands for the irradiation time and $k$ for the degradation rate constant. Values of $k$ determined from the slope of the linear curves in Fig. 4b, range from $3.74\times10^{-4}$ (raw anatase) to $1.66\times10^{-2}$ min$^{-1}$ (MW-500). The MW reaction process enhances the photocatalytic degradation–reaction rate by two orders of magnitude. Fig. 4c shows that RhB molecules are also more readily adsorbed on the graphene coating obtained through this process (MW-500). This behaviour might be imputed to CH/π and/or π/π van der Waals interactions involving the aromatic rings or via electrostatic attractions [28] expected between RhB and graphene or C/TiO$_2$ composite [29]. Removal of ROC by toluene increases the adsorption capacity in the case of MW-500. Comparison of optical spectra (Fig. S16) reveals that the MW-assisted coated TiO$_2$ absorbs photons in the broader range from UV to visible and NIR regions as compared to other photocatalysts. The recombination rates of the photo-excited carriers are evaluated on the basis of photoluminescence of Anatase (Fig. S17). The two peaks at 389 nm (A$_1$) and 426 nm (A$_2$) are assigned to the recombination of a bulk exciton, and a shoulder at 540 nm (A$_4$) corresponds to the recombination of electrons with oxygen vacancy [30]. In addition, the peak at 492 nm (A$_3$; equivalent to 2.52 eV) might be partially attributed to defects accompanied with C doping, shows linearly increase in the peak area ratio as a function of C doping concentration (calculated by XPS spectra) in the synthesized TiO$_2$ (Fig. S18). This result is consistent with the bandgap narrowing by the localized energy levels formed about 0.60 eV above VBM in C doped TiO$_2$ [31]. Fig. S19 shows the comparison of the relative peak area ratio of A$_4$ between the pristine and the synthesized samples by MW and conventional methods. The electron exited in TiO$_x$ rapidly transfers into the graphene shell as efficient carrier separation, resulting in the decrease of the relative peak area ratio. This tendency can be seen more clearly in the synthesized sample through MW-assisted reaction might due to the highly uniform formation of graphene shell on the particle. As the impurity level would act as a recombination centre between excited carriers, this results in the deterioration of photocatalytic activity. The CB edge of graphene with oxygen incorporations consists mainly of anti-bonding π* orbitals, situated at a higher energy level than that required for the reduction of the dissolved oxygen in water by electrons ($O_2^-$ (aq)/$O_2$) [32]. The VB edge of the graphite oxide formed by the $O_{2p}$ orbital is located at -5.9 eV, from the vacuum level [32]. Thus, photo-excited holes in $TiO_x$ can easily be transferred to the VB edge of the oxidative graphene. In contrast, the formation of both $Ti^{3+}$ species and oxygen vacancies below a minimum CB [33] induces the localised energy levels, that act as stronger electron acceptors, leading to an excellent spatial photo-carrier separation as shown in Fig. 4d in addition to uniform electronic contact in the interface of core/shell $TiO_2$/$TiO_x$, fully covered by the graphene shell. For carbon-coated photocatalysts, the thickness of the carbon shell is a key factor in improving the photocatalytic performance because irradiated light is strongly weakened by scattering and absorption in thicker shells before arriving at a photocatalyst's particle surface [34]. In our case, the homogeneous pore-free graphene coating would not garner these negative effects of light absorption in $TiO_x$/$TiO_2$. By improving both the light absorption and the organic molecular adsorption, the MW assisted graphene coating leads to higher photocatalytic performances.

As conclusion, we propose a novel chemical reaction pathway for rapid build-up of the single-nano sized graphene/$TiO_x$ multi-shells on $TiO_2$ by MW-assisted concomitant reaction on the metal oxide / polymer interface in inert atmosphere conditions. The concomitant reaction includes rapid polymer chain cleavage accompanying oxygen abstraction from the neighbouring oxide particle surface, resulting in faster growth of graphene shells with oxygen incorporation on the modified $TiO_2$ following the direct formation of small molecules with C=C and aromatic rings and dissociative adsorption on the formed $TiO_x$. This mechanism seems supported by the action of the MW electric field and provides a new insight for composite materials design using organic/ceramics interface, which is not taking place in resistive furnace heating. Furthermore, the visible-light photocatalytic performance of these $TiO_2$ particles with homo-/ hetero-junction is enhanced by the synergetic effects between the following three factors; significant enhancement of the light absorption capacity, the adsorption capacity of organic molecular, and the efficiency of excited carriers spatial separation.

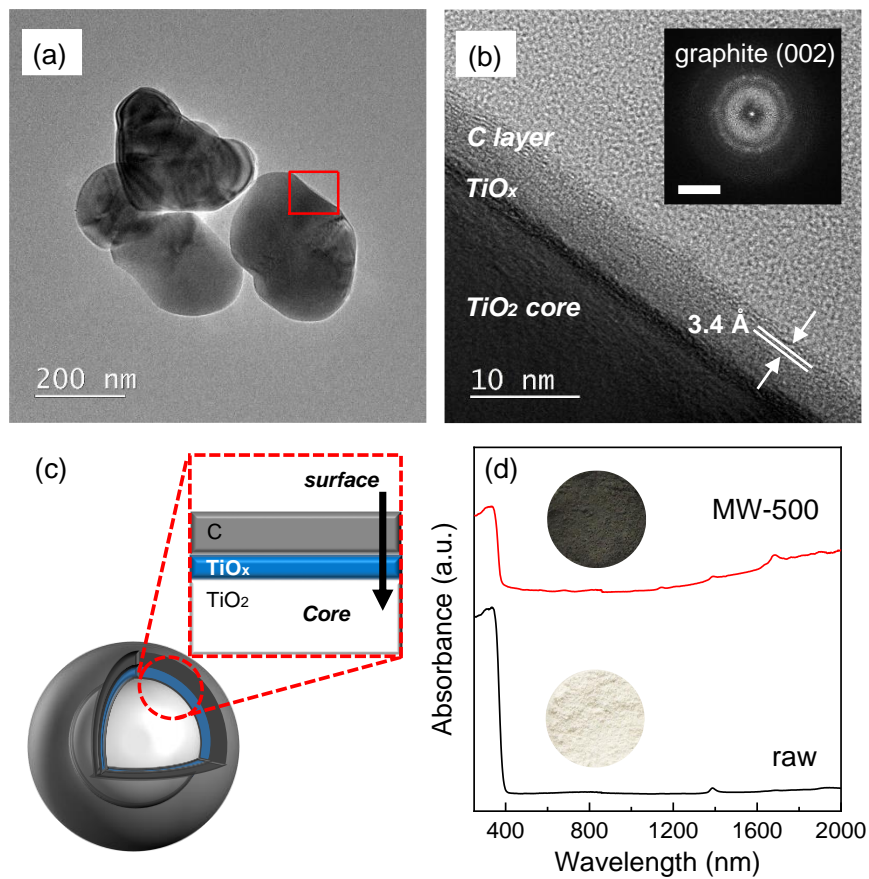

**Figure 1.** Representative morphology and optical property of the composite particles obtained by MW-assisted reaction. (a) low magnification micrograph and (b) HR-TEM of the red dash square area showing the discrete multi-shelled structure of the particle. The SAED pattern (inset) obtained for the outermost shell suggests a graphitic interplanar distance (white scale bar: 2 nm$^{-1}$). (c) Diagrammatic representation of the $TiO_2/TiO_{2-x}/C$ particle structure. (d) comparative UV-vis-NIR absorption spectra, and associated apparent colours (insets) of the composite before and after MW irradiation.

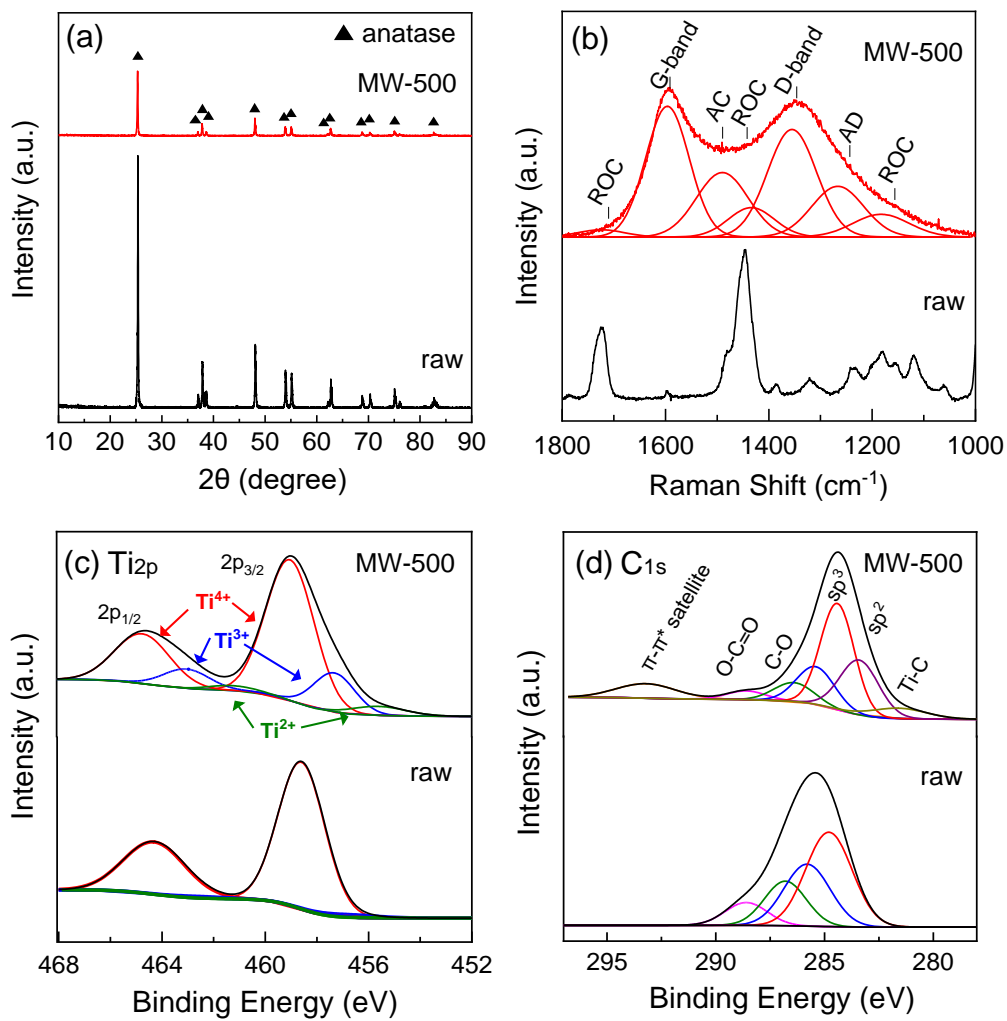

**Figure 2.** Chemical structure characterisation for the raw powder mixture and the obtained particle. (a) XRD patterns. (b) Raman spectra. (c) XPS Ti$_{2p}$ and (d) XPS C$_{1s}$ spectra.

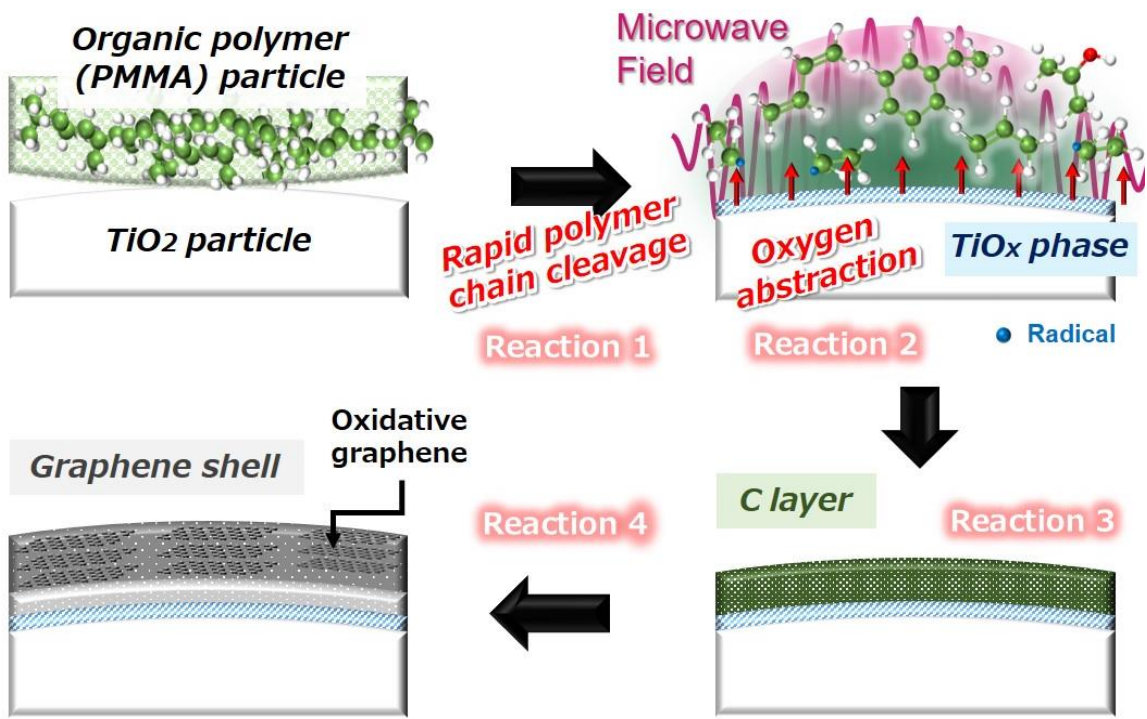

**Figure 3.** Schematic illustration for the formation mechanism of ultra-thin graphene-TiO$_x$ multi-shelled TiO$_2$ via the MW-assisted reaction. (Reaction 1) Polymer chains dissociation in MW electric field, leading to active low molecular fragments containing unsaturated C=C bonds and aromatic rings. (Reaction 2) Reduction of TiO$_2$ surface via oxygen abstraction by the radical species in the small molecular fragments. (Reaction 3) Dissociative adsorption of the fragments on the formed TiO$_x$. (Reaction 4) Acceleration of conversion into oxidative graphene.

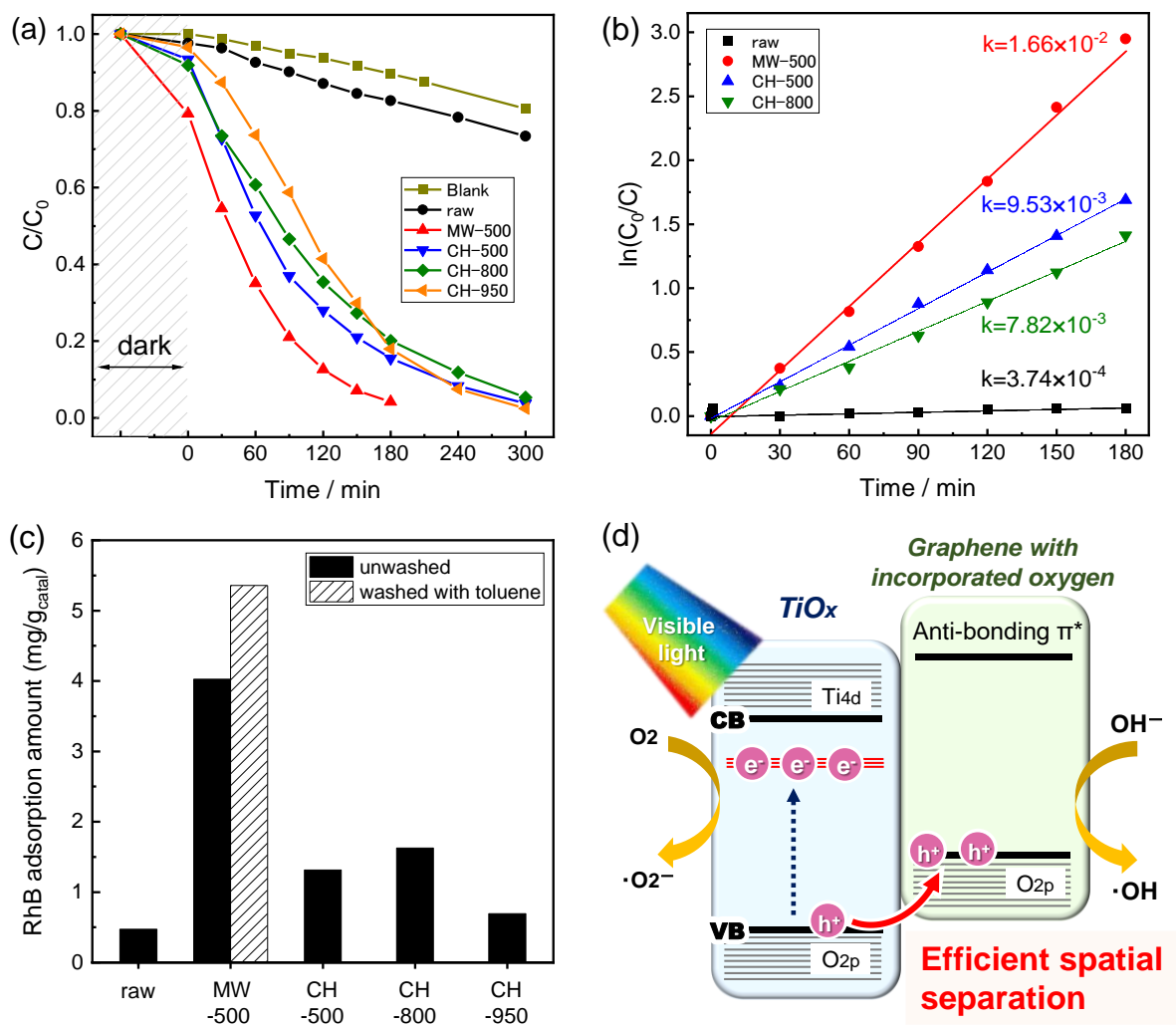

**Figure 4.** Comparison of photocatalytic performance (MW-X = MW heating; CH-X = conventional heating). (a) Photocatalytic activity for photo-degradation of RhB under 10 mW cm$^{-2}$ visible-light irradiation (385–740 nm) and (b) corresponding degradation rate. (c) Surface adsorption capacity of the photocatalysts for RhB. The black and striped columns represent the RhB adsorption amount (mg/g$_{catal}$) of the unwashed and washed sample by toluene, respectively. (d) Schematic illustration of the band structure of the multi-shelled TiO$_2$.